\documentclass[11pt,superscriptaddress, titlepage]{revtex4-1}  
\usepackage{amsmath}  
\usepackage{graphicx}  
\usepackage{dcolumn}   
\usepackage{bm}        
\usepackage{amssymb}   
\usepackage[colorlinks,hyperindex]{hyperref}
\usepackage{hyperref}
\usepackage{mathptmx}
\usepackage[T1]{fontenc}
\usepackage{lmodern}
\usepackage{titlesec}
\usepackage{fancyhdr}
\usepackage{lastpage}
\usepackage{lipsum}
\usepackage{array,multirow}
\fancypagestyle{report}%
\hypersetup
	{
		colorlinks,%
		citecolor=green,%
		linkcolor=black,%
		urlcolor=black,%
	}

\renewcommand{\thesection}{\Roman{section}}
\begin{document}
\widetext
\title{\Large{Type-III Seesaw fermionic triplets at the International Linear Collider}}
\author{Deepanjali Goswami}
\thanks{g.deepanjali@iitg.ernet.in/goswamideepanjali@gmail.com}
\affiliation{Department of Physics, Indian Institute of Technology Guwahati, Guwahati-781039, India.}
\author{P. Poulose}
\thanks{poulose@iitg.ernet.in}
\affiliation{Department of Physics, Indian Institute of Technology Guwahati, Guwahati-781039, India.}

\begin{abstract}
The signature of heavy fermionic triplets belonging to Type III seesaw at the International Linear Collider (ILC) is probed.
Presence of charged fermionic triplets upto a mass of about $ 750$ GeV could be established through single production at a 1 TeV ILC with moderate luminosity of 300 fb$^{-1}$, assuming a fermion triplet-electron mixing of about 0.05. Unlike the case of LHC, the production process is highly sensitive to the mixing, making the process interesting. The single production of neutral triplet is found to be somewhat harder, considering the large SM background present. Pair production of triplets of mass 500 GeV considered at 2 TeV centre of mass energy presents convenient ways to study different mixing scenarios. The production process is sensitive to $V_e$. The pair production along with information regarding single production would be able to identify the mixing scenarios. 
\end{abstract}


\maketitle
\newpage
\section{Introduction}
\label{intro}
\noindent The Standard Model (SM) of particle physics presumes that the neutrinos are massless, and therfore does not provide any way to accommodate their tiny mass, which is now established by the experiments \cite{neutrino mass}. Understanding the mechanism to generate neutrino mass is one of the main issues of particle physics today.
Among various explanations offered for this, the seesaw mechanism seems most plausible way. In this mechanism the lightness of neutrino is connected to a new large mass scale (M), usually brought in as the mass of a heavy partner. Mainly three different types of seesaw mechanism are proposed, and studied in the literature. The Type I seesaw model \cite{typeI} has a gauge singlet right-handed neutrino field in addition to the SM field. In Type II seesaw model \cite{typeII} a scalar  $SU(2)_{L}$ triplet field with hypercharge $Y=2$ is added, and in Type III seesaw model \cite{typeIII, Li:2009mw, typeIIImodel}  a fermionic triplet field with $Y= 0$ is added.  

The neutrino experiments can only probe the combination $y/M$ of the Yukawa couplings ($y$) and the large mass scale associated with the new particles in the spectrum ($M$), but does not give any independent information about $M$. If $y\sim\mathcal{O}(1)$ then $M\sim 10^{15}$ GeV for the mass of the neutrinos, $m_{\nu}$ in the eV range. This scale is far beyond the reach of any conceivable accelarator. However, this scale can in principle be as low as $M\sim$ 100 GeV to 1 TeV, which  can be easily tested at the Large Hadron Collider(LHC) and the proposed International Linear Collider (ILC). This will also lead to interesting phenomenological signatures at these machines. 
We shall consider the Type III seesaw model, in  which the seesaw mechanism is realised through a fermionic triplet. The study of the phenomenology of Type-III seesaw model in the neutrino sector is carried out  by, for example, Ref.\cite{Ahn:2011pq}.  This model predicts three additional heavy fermions, one of which is neutral. The presence of such heavy fermion at LHC has been searched for \cite{ATLAS-CONF,Aad:2015cxa}, and signatures are studied phenomenologically \cite{Biggio:2011ja}, through single and pair production of the heavy fermions.  Note that, these heavy fermions mix with the SM leptons. However, the production at LHC is through gauge couplings, and therfore not sensitive to the mixing. The decay, on the other hand, could be sensitive to these mixings. But, as discussed in the following sections, it is usually difficult to probe the mixing at LHC. Proposing an electron-proton collider, single production of the charged and neutral heavy fermions in the expected LHeC is studied in Ref.~\cite{Liang:2010gm}. Here, of course, the production itself could be sensitive to the mixing, owing to the electron in the initial state. For phenomenological studies in the context of ILC involving Higgs bosons and the fermion triplets are presented in Ref. \cite{ILCpheno}. 

In the present work we focus our attention on the production of fermion triplets at the ILC and the study of identification of these triplets over the SM backgrounds. The advantages of ILC, projected as a precision machine \cite{ILC}, include fixed centre of mass energy of the initial states, and availability of beam polarisation \cite{ILCpolarisation}. In the present case, ILC, cite
 unlike the LHC has the advantage that the production itself could be sensitive to the mixing between the heavy fermions and the SM leptons. We shall see how  the mixing can be probed through the processes studied here.

The organization of the paper is as follows. In section \ref{typeiii} we will give the information of the Type-III seesaw model. In section \ref{process} we will describe the processes under study and we will conclude the work in section \ref{conclusion}.

\section{Type-III seesaw model}
\label{typeiii}
The Type III seesaw model contains an $SU(2)$ triplet fermion field, denoted here as $\Sigma$, in addition to the SM fields.
The Lagrangian corresponding to the  model is given by \cite{Biggio:2011ja}

\begin{eqnarray}
\mathcal{L} &=& \mathcal{L}_{SM} + \mathcal{L}_{\Sigma},\nonumber
\end{eqnarray}
with 
\begin{eqnarray}
\mathcal{L}_{\Sigma}&=&
\textrm{Tr\ensuremath{\left[\bar{\Sigma\,}\slash\!\!\!\!\!\!D\Sigma\right]}}-
\frac{1}{2}Tr \left[\overline{\Sigma}M_{\Sigma}\Sigma^{c}+\bar{\Sigma}^{c}M_{\Sigma}^*\Sigma\right]
- \tilde{\phi}^{\dagger}\bar{\Sigma}\sqrt{2}Y_{\Sigma}L
- \bar{L}\sqrt{2}Y_{\Sigma}^{\dagger}\Sigma\tilde{\phi},
\end{eqnarray}
\noindent where $\mathit{M}_{\Sigma}$ is the mass matrix of the triplet
and $\mathit{Y}$$_{\Sigma}$ is the Yukawa coupling matrix, $\mathit{L}\equiv(l,v)^{T}$ is the SM left-handed doublet lepton field, $\phi$$\equiv$($\phi^{+},$$\phi^{0})^{T}$$\equiv$($\phi^{+}$,($\mathit{v}$+$\mathit{H}+$ $\mathit{i}$$\eta$)/$\sqrt{2}$ )$^{T}$
is the Higgs field, and $\tilde{\phi}$ = $\mathit{i}$ $\tau_{2}$ $\phi^{*}$. The fermion triplet $\Sigma$ is explicitly given by
\begin{equation}
\Sigma = \begin{pmatrix}\Sigma^{0}/\sqrt{2} & \Sigma^{+}\\
\Sigma^{-} & -\Sigma^{0}/\sqrt{2}
\end{pmatrix}
\end{equation}
and its conjugate $\Sigma^{c}\equiv\mathit{C}\bar{\Sigma}^{T}$.

A Dirac spinor \(
\Psi \equiv \Sigma^{+c}_{R} + \Sigma^{-}_{R}\)  
is defined to conveniently express the mixing of the SM charged leptons with the triplets.
The neutral component of the fermionic triplet, which mixes with neutrinos can be left as the two component spinors, as they have only two degree of freedom. The corresponding  Lagrangian 

\begin{eqnarray}
\mathcal{L}_{\Sigma} & = & \overline{\Psi}i \slash\!\!\!\!{\partial}\Psi + \overline{\Sigma}^{0}_{R}i \slash\!\!\!\!{\partial}\Sigma^{0}_{R} - \overline{\Psi} M_{\Sigma}\Psi - \left(\frac{1}{2}\overline{\Sigma}^{0}_{R}~M_{\Sigma}~\Sigma^{0c}_{R} + h.c \right)  + g\left( W^{+}_{\mu} \overline{\Sigma}^{0}_{R}\gamma_{\mu} P_{R} \Psi + W^{+}_{\mu}\overline{\Sigma}^{0c}_{R} \gamma_{\mu}P_{L}\Psi + h.c.\right) \nonumber \\&& - gW^{3}_{\mu}\overline{\Psi}\gamma_{\mu} \Psi - \left(\phi^{0}\overline{\Sigma}^{0}_{R}Y_{\Sigma}\nu_{L} + \sqrt{2}\phi^{0}\overline{\Psi}Y_{\Sigma}\ell_{L} + \phi^{+}\overline{\Sigma}^{0}_{R}Y_{\Sigma}\ell_{L} -      \sqrt{2}\phi^{+}\overline{\nu}^{c}_{L}Y^{T}_{\Sigma}\Psi + h.c \right)  
\end{eqnarray}
After diagonalising the mass matrices of the charged and the neutral sectors, the Lagrangian in the physical basis is given in the Appendix \ref{appendix} \cite{typeIIImodel}

Note that we need atleast two triplets to produce two non-vanishing mass-differences, as is observed. For the simplicity, we have considered only one triplet in the physical spectrum in the  GeV scale where the Yukawa coupling matrix can be written in $1\times3$ vector representation as
\(
Y_{\Sigma} = \begin{pmatrix}
Y_{\Sigma_{e}} & Y_{\Sigma_{\mu}} & Y_{\Sigma_{\tau}} \end{pmatrix}.   
\)
The assumption of real parameters allows to write all the couplings in terms of the mixing parameters, $V_{\alpha}$, where $\alpha$ labels the couplings to each of the $e$, $\mu$ and $\tau$ lepton generation as
\begin{equation}
V_{\alpha} = \frac{v}{\sqrt{2}} M^{-1}_{\Sigma}Y_{\Sigma_{\alpha}}.
\label{valpha}
\end{equation}
Defining the product \(\vert V_{\alpha}V_{\beta} \vert = \frac{\upsilon^{2}}{2} \vert Y^{\dagger}M^{-2}Y \vert_{\alpha\beta} \), bounds on the mixing parameters, obtained from the flavour changing rare decay $\mu \rightarrow e\gamma$, $\tau \rightarrow \mu\gamma$ and $\tau \rightarrow e\gamma$ in the presence of one or more triplets of fermions, are given by \cite{typeIIImodel, Abada:2007ux, delAguila:2008pw, Biggio:2011ja}
\begin{eqnarray}
&&\vert V_{e} \vert < 0.055,~~ 
\vert V_{\mu} \vert < 0.063, ~~
\vert V_{\tau} \vert < 0.63 \\
&&\vert V_{e} V_{\mu} \vert < 1.7 \cdot 10^{-7},~~
\vert V_{e}V_{\tau} \vert < 4.2 \cdot 10^{-4},~~
\vert V_{\mu}V_{\tau} \vert < 4.9 \cdot 10^{-4}
\end{eqnarray}
The triplets decay to SM leptons, gauge bosons and the Higgs boson. The decay widths of different channels are given by \cite{Franceschini:2008pz}

\begin{eqnarray}
\Gamma(\Sigma^{0} \rightarrow l^{-}_{\alpha}W^{+}) &=& \Gamma(\Sigma^{0} \rightarrow l_{\alpha}^{+}W^{-}) = \frac{g^{2}}{64 \pi} \vert V_{\alpha} \vert^{2} \frac{M^{3}_{\Sigma}}{M^{2}_{W}} \Big( 1- \frac{M^{2}_{W}}{M^{2}_{\Sigma}}\Big)^{2}  \Big(1+ 2\frac{M^{2}_{W}}{M^{2}_{\Sigma}}\Big)\\
\sum\limits_{l} \Gamma(\Sigma^{0} \rightarrow \nu_{l}Z) &=& \frac{g^{2}}{64\pi c^{2}_{W}} \sum\limits_{\alpha} \vert V_{\alpha} \vert^{2} \frac{M_{\Sigma}^{2}}{M_{Z}^{2}} \Big(1 - \frac{M^{2}_{Z}}{M^{2}_{\Sigma}}\Big)^{2} \Big(1 + 2\frac{M^{2}_{Z}}{M^{2}_{\Sigma}}\Big)\\
\sum\limits_{l} \Gamma(\Sigma^{0}\rightarrow \nu_{l}H)& =& \frac{g^{2}}{64\pi} \sum\limits_{\alpha}\vert V_{\alpha} \vert^{2} \frac{M^{3}_{\Sigma}}{M^{2}_{\Sigma}}\Big(1 - \frac{M_{H}^{2}}{M^{2}_{\Sigma}}\Big)^{2}\\
\hspace{2cm}\sum\limits_{l} \Gamma(\Sigma^{+} \rightarrow \nu_{l}W^{+}) &=& \frac{g^{2}}{32\pi} \sum\limits_{\alpha} \vert V_{\alpha} \vert^{2} \frac{M_{\Sigma}^{3}}{M^{2}_{W}} \Big(1 - \frac{M^{2}_{W}}{M^{2}_{\Sigma}} \Big)^{2} \Big(1 + 2\frac{M^{2}_{W}}{M^{2}_{\Sigma}}\Big)\\
\Gamma(\Sigma^{+} \rightarrow l^{+}_{\alpha}Z) &=& \frac{g^{2}}{64\pi c^{2}_{W}} 
\vert V_{\alpha} \vert^{2} \frac{M^{3}_{\Sigma}}{M^{2}_{Z}} \Big(1 - \frac{M^{2}_{Z}}{M^{2}_{\Sigma}}\Big)^{2} \Big(1 + 2\frac{M^{2}_{Z}}{M^{2}_{\Sigma}}\Big)\\
\Gamma(\Sigma^{+} \rightarrow l^{+}_{\alpha}H) &=& \frac{g^{2}}{64\pi} \vert V_{\alpha} \vert^{2}\frac{M_{\Sigma}^{3}}{M^{2}_{W}}\Big(1 - \frac{M^{2}_{H}}{M^{2}_{\Sigma}}\Big)^{2}
\label{eq:decays}
\end{eqnarray}

Direct searches by LHC \cite{Aad:2015cxa} exclude the mass of the heavy triplets below 320- 540 GeV at the 95$\%$ Confidence Level, depending on different mixing scenarios considered. As mentioned above, it is hard to obtain information regarding mixing parameters at LHC. Firstly, the production mechanisms involve only gauge couplings of the triplets, and therefore are blind to mixing. The decay widths, on the other hand has strong dependence on the mixings. However, in the total cross section, considered as product of production cross section and branching ratio of the decay channel considered, this dependence is mellowed down. For example, assuming only one mixing is present, the dependence will be cancelled away in the branching ratio. The advantage of ILC in this regard is evident, as the production mechanism itself could depend on the electron-triplet mixing parametrised through$V_e$. In the next section we shall consider the single and pair production of the triplets at ILC.

\section{Processes considered for  study}
\label{process}
At ILC $\Sigma^+\Sigma^-$ and $\Sigma^0\Sigma^0$ pair production, as well as charged and neutral single production of the triplets in association with the SM leptons can be studied. Complying with the recent direct limits, we shall consider $M_\Sigma=500$ GeV or above for our numerical studies. For pair production we need to go to centre of mass energies of 1 TeV or above. At around 1TeV the cross section is negligible owing to little phase space available. We have therefore considered a 2 TeV ILC for our study. On the other hand, single production could be searched for at a 1 TeV collider.  As discussed later, charged triplets of mass close to about a TeV could be possibly identified above the SM background at a 1 TeV ILC. The case of neutral triplet is less promising, considering the much larger background associated with this.

\subsection{Single production of $\Sigma^{0}$ and $\Sigma^\pm$}

\begin{figure}[h]
\includegraphics[width = 5cm]{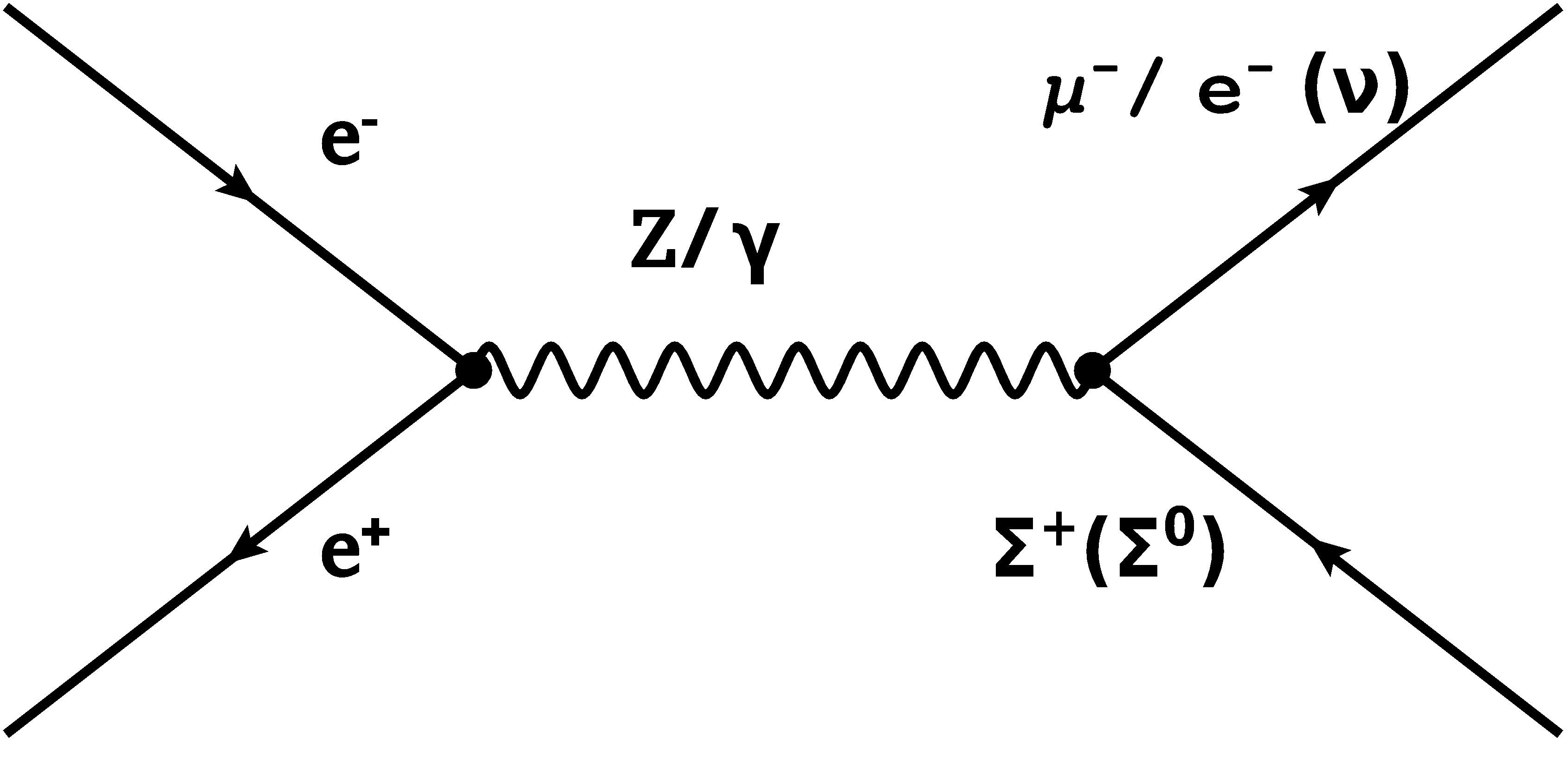}\hskip 10mm
\includegraphics[width=3cm]{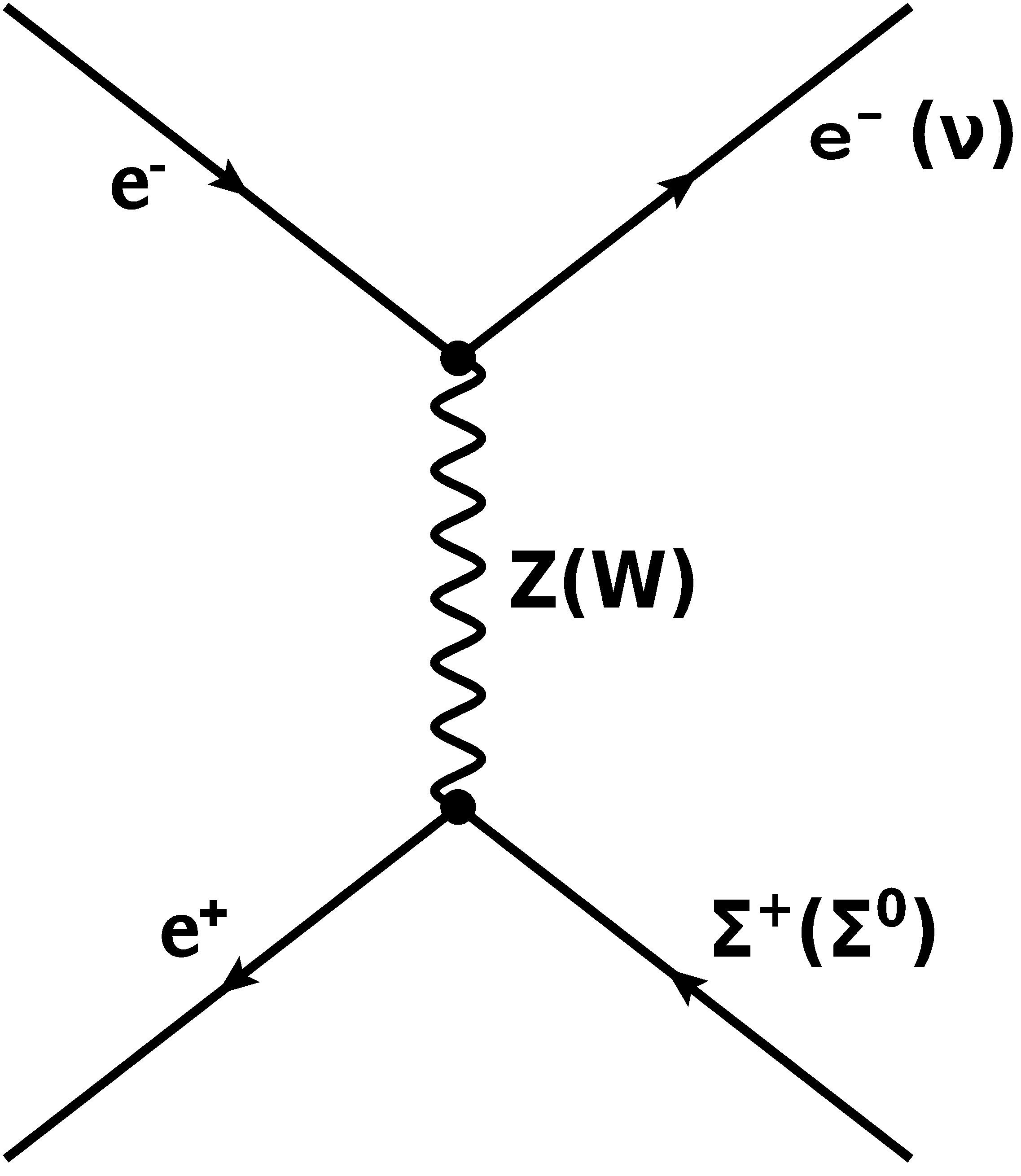}
\caption{Feynman diagrams contributing to the process $e^{+} e^{-} \rightarrow \Sigma^{+} \ell^{-}$ ($\Sigma^{0}\nu$) at the ILC}
\label{fig:fd-single}
\end{figure}

Considering the single production of neutral and charged components of the fermion triplet along with a neutrino or lepton, respectively is a suitable candidate to probe the mixing. The Feynman diagrams involve an $s$-channel exchange of gauge bosons, and, when $V_e\ne0$, an additional $t$-channel, as shown in Fig.~\ref{fig:fd-single}. 

\begin{table}[h]
\begin{center}
\begin{tabular}{|c|l|r|r|}
\hline
&&\multicolumn{2}{|c|}{$\sigma$ in fb}\\\cline{3-4}
 $\sqrt{s}$&Process  &$V_{e}=0.05,  $& $V_{\mu}= 0.05,$\\
 &&$V_{\mu}=V_{\tau}=0$&$V_{e}=V_{\tau}=0$ \\ 
 \hline
 \hline
$1$ TeV& $\sigma(e^{-} e^{+} \rightarrow \Sigma^{0}\nu$)  & 186 & 0.078 \\ 
& $\sigma(e^{-} e^{+} \rightarrow \Sigma^{+}\ell^{-}$)  & 31 & 0.078 \\\hline
$2$ TeV& $\sigma(e^{-} e^{+} \rightarrow \Sigma^{-} \Sigma^{+})$  & 48.03 & 55.6 \\
& $\sigma(e^{-} e^{+} \rightarrow \Sigma^{0} \Sigma^{0})$  & 0.47 & $2.5\times 10^-5$ \\ \hline
\end{tabular}
\caption{Cross-section for the pair and single production of charged and neutral fermion  with different mixing angles, at the ILC with CM energy ($\sqrt{s}$) of 2 TeV and 1 TeV, respectively, with $M_{\Sigma}$  of 500 GeV. }
\label{table:cross-section}
\end{center}
\end{table}
Table~\ref{table:cross-section} gives the production cross section at a 1 TeV ILC for two different mixing scenarios, involving $V_e\ne0$, and $V_e=0$.  The cross section is proportional to $\sum\limits_\alpha |V_\alpha|^2$ in case of neutral triplet production, and to the individual $|V_\alpha|^2$ in the case of charged triplet production. The $s$-channel contribution is negligible, as it falls down with $\sqrt{s}$. The $t$-channel gives substantial contribution when $V_e\ne 0$, with 186 fb cross section for neutral triplet with mass $M_\Sigma=500$ GeV, and 31 fb cross section for charged triplet of the same mass. The production associated with muon is negligible. Thus, while the presence of $\Sigma^\pm$ along with $e^\mp$ will clearly indicate the presence of $V_e$, the presence or absence of $V_{\mu,~\tau}$ would not be established as easily through the single triplet production.

The decay branching ratios (BR) of the triplets to channels specified in Eq.~\ref{eq:decays} are given in Table~\ref{table:BR} for $M_\Sigma=500$ GeV.  Of the charged triplets, about two-third decay to $W\nu$, and one-third to $Zl$, with a tiny 1\% decaying to $Hl$.  The neutral triplets decay 49\% of the time each to $Wl$ and $Z\nu$, and about 2\% to $H\nu$. In our further analyses we shall neglect the decay to the Higgs bosons. With these, the $\Sigma^+\Sigma^-$ pair production leads to the following three distinct signals.

\begin{table}[h]
\begin{center}
\begin{tabular}{|l|l|l|l|}
\hline
 \multicolumn{2}{|c|}{$\Sigma^\pm$}&\multicolumn{2}{|c|}{$\Sigma^0$} \\ \cline{1-4}
Channel&BR in $\%$ &Channel&BR in $\%$ \\ \cline{1-4}
$\Sigma^{+} \rightarrow W^{+}\nu$&65.91 & $\Sigma^{0} \rightarrow W^{+}\ell^{-}$ &49.15\\
$\Sigma^{+} \rightarrow Z\ell^{+}$ &32.91 & $\Sigma^{0} \rightarrow Z\nu$ &49.09\\
$\Sigma^{+} \rightarrow H\ell^{+}$ &1.17  & $\Sigma^{0} \rightarrow H\nu$ &1.75 \\ \cline{1-4}
\end{tabular}
\caption{Branching ratio of the charged and neutral triplet with mass, $M_\Sigma=500$ GeV.  }
\label{table:BR}
\end{center}
\end{table}

Different final states arising from the above considerations are listed in Table \ref{table:finalstate-nusingle}, along with possible SM backgrounds. The $WW$ background in the case of final states involving at least one lepton, and the continuum $2j$ background in the case of $2j+$MET are large. While the $q\bar q$ background can be contained demanding reasonable missing transverse energy, the $WW$ background is hard to eliminate, as it contains missing energy, like the signal. At the same time, detailed analysis of the missing energy and angular distributions should be able to reduce the background considerably, as the topology of both the processes are distinct. 

\begin{table}[h]
\begin{center}
\begin{tabular}{|l|l|l|l|l|}
\hline
Final State & Signal & \multicolumn{2}{c|}{$\sigma\times$ {BR} in fb} & Background\\\cline{3-4}
&  &$V_{e}=0.05,  $& $V_{e}= 0,$&($\sigma\times$ BR in fb) \\
 &&$V_{\mu}=V_{\tau}=0$&$V_{\mu~ {\rm or}~\tau}\ne 0$& \\ 
\hline
\hline
2$j$ + $\ell$+ MET & $W^{+} \ell^{-}\nu$ & 61.6 & 0.026 & WWZ(0.86), WW(193.5) \\
2$\ell$ + MET & $ W^{+} \ell^{-}\nu $ & 9.82 & 0.004 &  WWZ(0.1), WW(31.0) \\ 
\hline
2$\ell$ + MET  & Z$\nu \nu$ & 3.07 & 0.001 &   WWZ(0.1), WW(31.0) \\
2$j$ + MET  & Z$\nu \nu$& 63.83 & 0.027 &  $q \bar{q}$ (347) \\
\hline
\end{tabular}
\caption{Signal and corresponding background for chosen final states for the signal  process $e^{-} e^{+} \rightarrow \Sigma^{0} \nu$ for $M_{\Sigma^{0}}$ = 500 GeV and CM energy of 1000 GeV.}
\label{table:finalstate-nusingle}
\end{center}
\end{table}

\begin{figure}
\includegraphics[width=9cm]{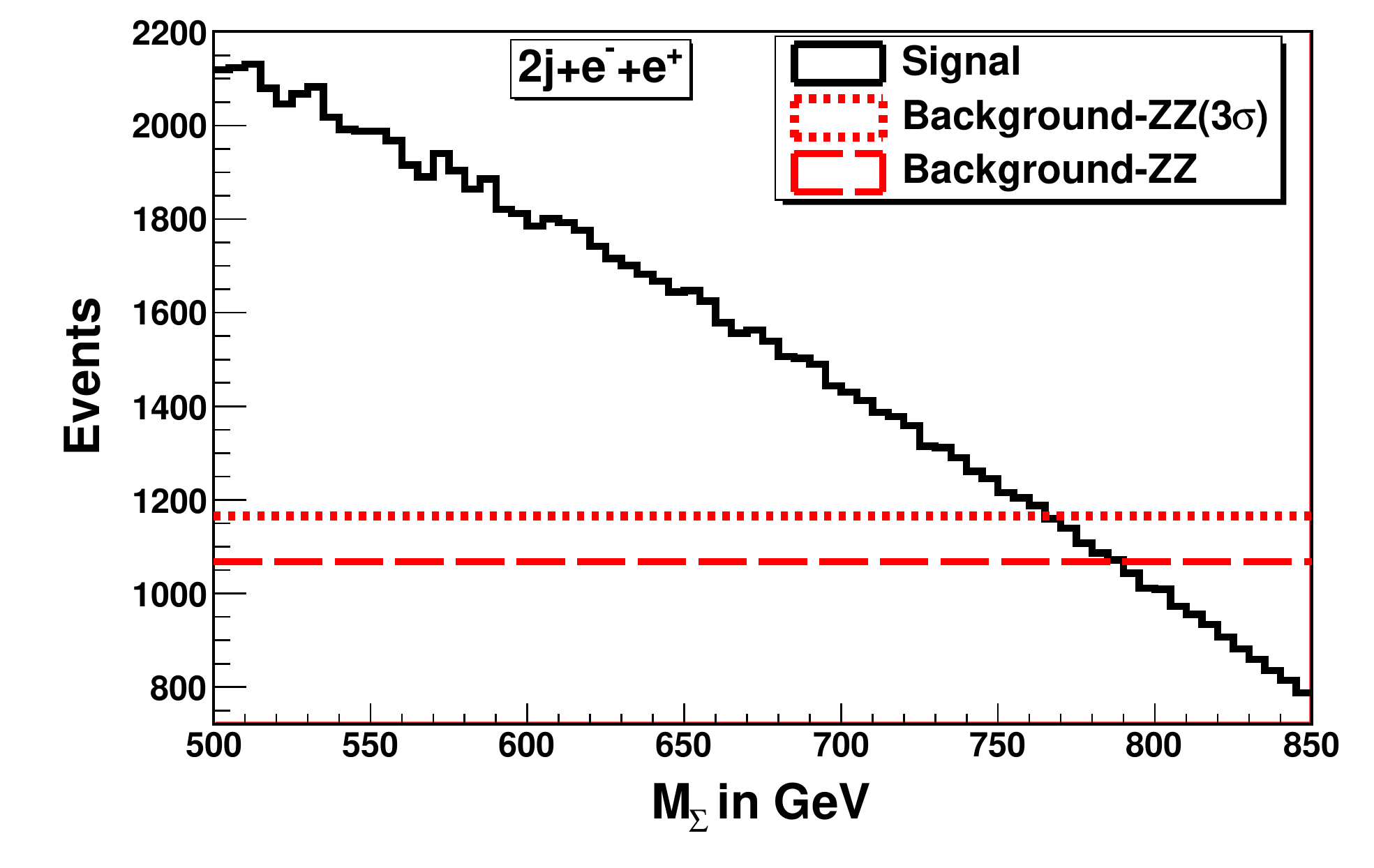}
\caption{Mass of Type-III seesaw fermion($\Sigma$) in GeV vs. number of events for the final state 2j + $e^{+}e^{-}$ from the signal Z$\ell\ell$ in the process $e^{+} e^{-} \rightarrow \Sigma^{+} \ell^{-}$ (black line) with mixing angle $V_{e} = 0.05, V_{\mu} = V_{\tau} = 0$. Red dashed and dotted lines represent the number of events produced by the ZZ background with CM energy of 1 TeV and integrated luminosity of 300 fb$^{-1}$ at ILC and its $3\sigma$ value, respectively. }
\label{fig:masslimit}
\end{figure}

Coming to the charged triplet production in association with charged SM leptons, possible final states along with corresponding SM backgrounds are given in Table~\ref{table:finalstate-chsingle}. Again, the $WW$ background is somewhat problematic in case of $2j+e+$MET and $e^+e^-+$MET. One may need to reduce the background with the help of kinematic distributions, in order to make use of these final states. On the other hand, the $2j+e^++e^-$ and $4l$ final states have relatively smaller backgrounds to worry about. The former one has sufficient cross section to study at moderate luminosities. We consider this channel to study the mass limit that could be probed at a 1 TeV ILC with integrated luminosity of 300 fb$^{-1}$. In Fig.~\ref{fig:masslimit} the number of events corresponding to this channel is plotted against $M_\Sigma$. The red-dash line corresponds to the SM background from $ZZ$ production, and the red- dot line is its $3\sigma$ limit. Assuming 100\% efficiency, this means that one would be able to probe up to about $M_\Sigma\sim750$ GeV. The other channels can support this to improve the limit, expectedly moderately.

\begin{table}
\begin{center}
\begin{tabular}{|l|l|l|l|l|l|}
\hline
Final State & Signal & \multicolumn{2}{c|}{$\sigma\times$ {BR} in fb} & Background\\\cline{3-4}
&  &$V_{e}=0.05,  $& $V_{e}= 0,$&($\sigma\times$ BR in fb) \\
 &&$V_{\mu}=V_{\tau}=0$&$V_{\mu~ {\rm or}~\tau}\ne 0$& \\ 
\hline
\hline
2$j$ + $e^{-}$+ MET &$W \ell \nu$ &  13.7 & 0.035 &  WWZ(0.86),  WW(193.5) \\
$e^{-}$ $e^{+}$ + MET & $ W \ell\nu $ & 2.2 & 0.006 &  WWZ(0.13), WW(31.0) \\ 
\hline
2$e^{-} +2e^{+}$ & Z $\ell \ell$ &  0.34 &  0.0009 & ZZ(0.17) \\
 $e^{-}$ $e^{+}$ +MET &Z$\ell \ell$ & 2.03 & 0.005  & WWZ(0.1), WW(31.0) \\
 2$j$ + $e^{-}$ $e^{+}$ &Z $\ell \ell$ &  7.13 & 0.018 &  ZZ(3.56)\\
\hline
\end{tabular}
\caption{Signal and corresponding background for chosen final states for the signal  process $e^{-} e^{+} \rightarrow \Sigma^{+} \ell^{-}$, cross-sections are with $\sqrt{s}$ = 1000 GeV and $M_{\Sigma}$ = 500 GeV at ILC.}
\label{table:finalstate-chsingle}
\end{center}
\end{table}

We may also note that the use of polarized beams, as proposed in the ILC studies \cite{ILCpolarisation}, will help especially the neutral triplet production involving a $t$-channel $W$ exchange. Further, 
as noted earlier, it is difficult to probe mixing with the second and third generation leptons through single production at 1 TeV. On the other hand, a $\mu^{-}\mu^{+}$ collider of suitable energy will be helpful here.

\subsection{Pair production of $\Sigma^{+} \Sigma^{-}$ and $\Sigma^0\Sigma^0$}

\begin{figure}[h]
\includegraphics[height=2.9cm]{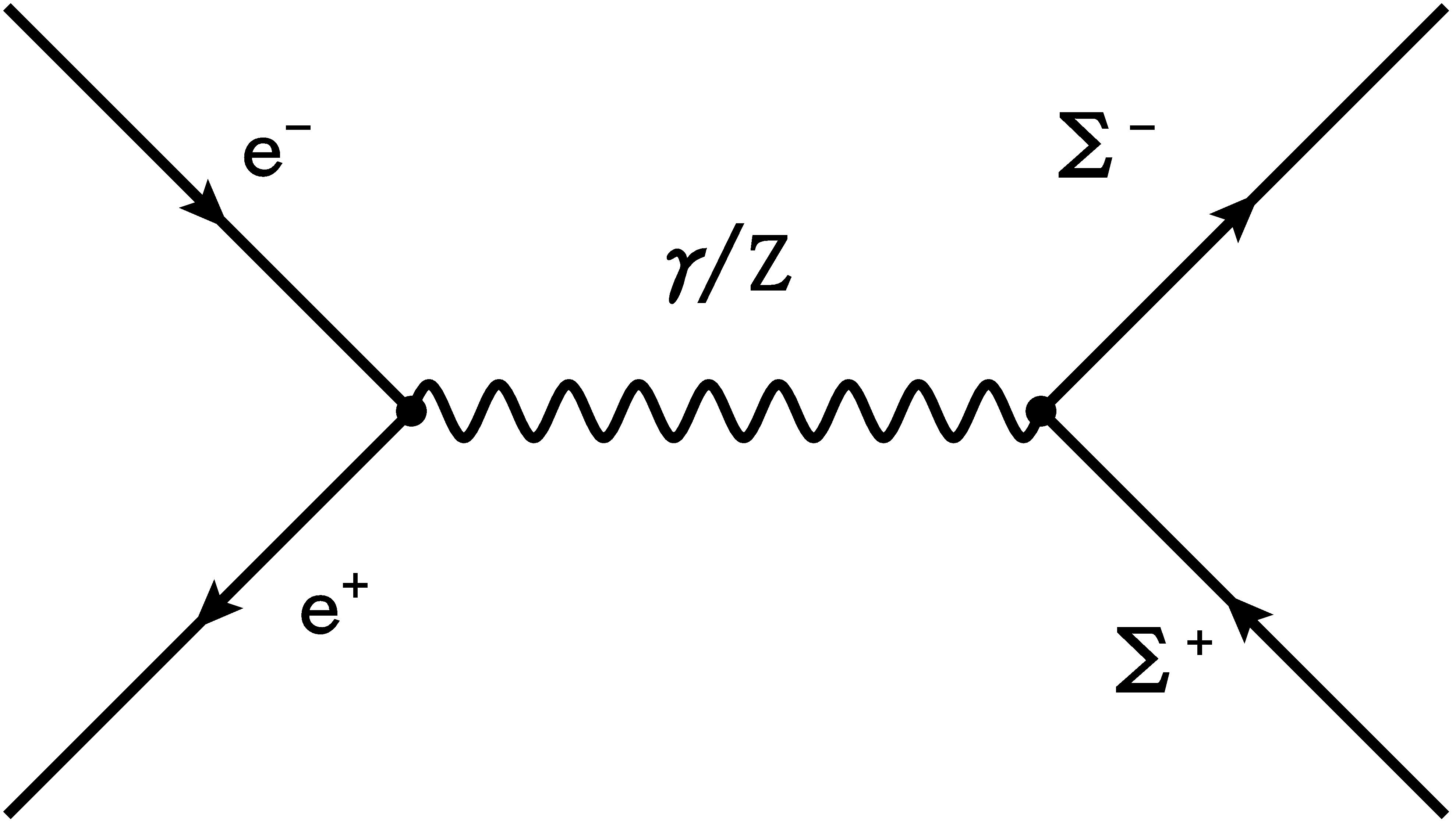} \hskip 10mm
\includegraphics[height=3.6cm]{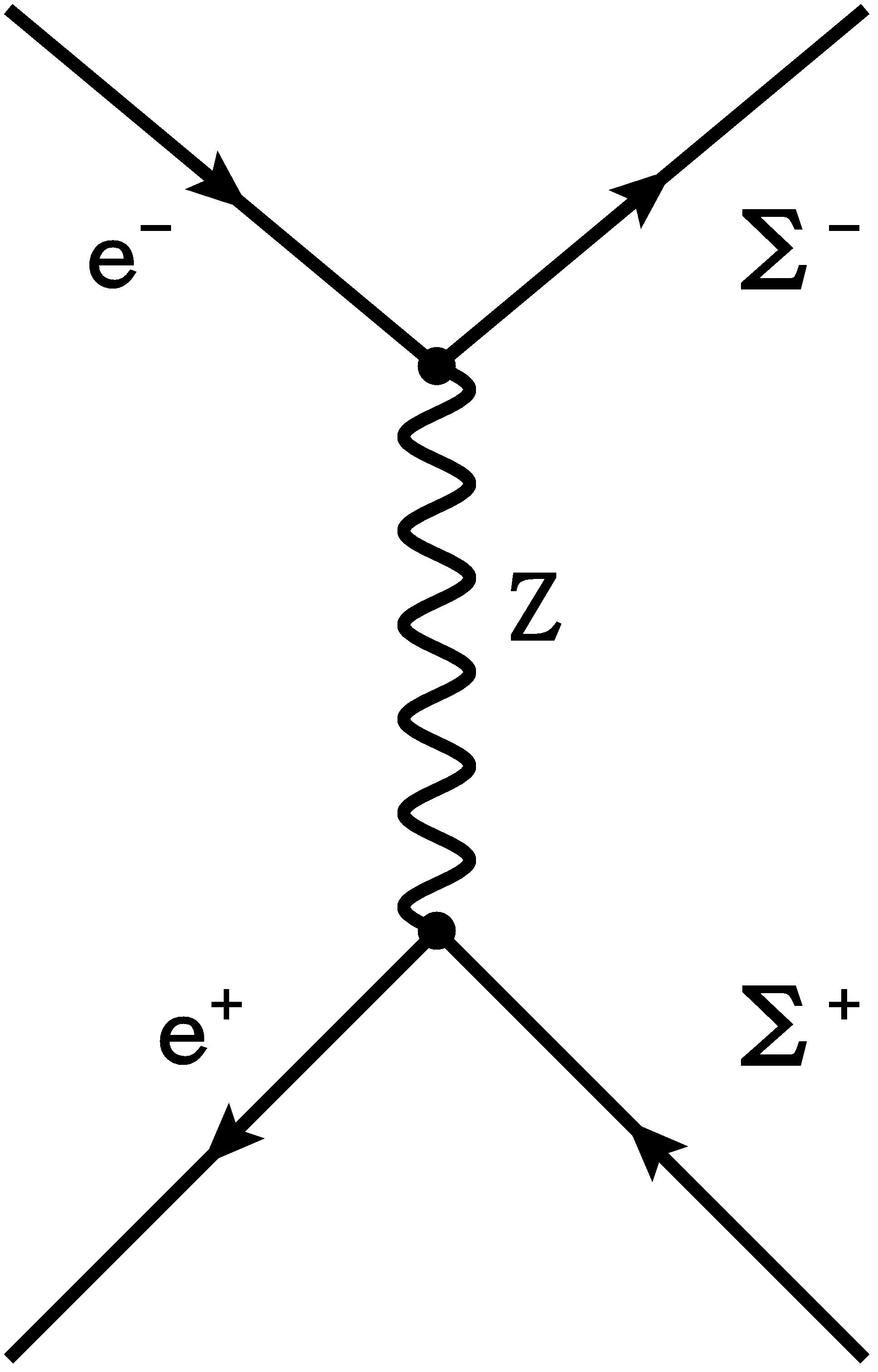}
\caption{Feynman diagrams contributing to the process $\mathit{e}^{+}e^{-}\rightarrow$ $\Sigma^{+}\Sigma^{-}$ in type-III seesaw model at the ILC.}
\label{fig:fd-pair}
\end{figure}

There are three possible Feynman diagrams for the pair 
production of $\Sigma$ at the ILC in the limit of vanishing electron mass which are shown in Fig.\ref{fig:fd-pair}. The $s$-channel mediated by the photon and $\mathit{Z}$ Boson
depends on all the three mixing parameters and thus proportional to the gauge coupling, while the $t$-channel depends on the value of $\mathit{V_{e}}$ only \cite{Garg:2014jva}.  Notice that the $t$-channel is absent at ILC, and therefore, production mechanism at LHC does not depend on mixing. This advantage of ILC could be exploited to see the sensitivity of the cross section on the mixing.

Typical mixing angles of $V_{e,\mu}=0.05$ is taken for illustration. The cross section for pair production of neutral triplet is very small, and therefore, we shall not discuss this any further. Considering pair production of the charged triplets, the case when $V_e=0$, the production cross section does not depend on the mixing, with the involvement only of the $s$-channel process.  This leads to a cross section of 55.6 fb (Table~\ref{table:cross-section}) With the $t$-channel switched on in the presence of non-zero $V_e$, this cross section is reduced, indicating negative interference.  For $V_e=0.05$ the cross section is reduced to close to 48 fb, which is a reduction of about 14\%. 

\begin{eqnarray}
\mathit{e^{+}e^{-}}&\rightarrow&\mathit{\Sigma^{+}\Sigma^{-}\rightarrow\mathit{W^{+}W^{-}\nu\nu}}\\
\mathit{e^{+}e^{-}}&\rightarrow&\mathit{\Sigma^{+}\Sigma^{-}\rightarrow ZZ \ell^+\ell^-}\\
\mathit{e^{+}e^{-}}&\rightarrow&\mathit{\Sigma^{+}\Sigma^{-}\rightarrow W^{+}\mathit{Z}\ell^{-}\nu}
\label{eq:ch-pair-signals}
\end{eqnarray}

With the $W$ and $Z$ decays included, the possible final states and corresponding cross sections are listed in Table~\ref{table:ch-finalstates}, along with the main SM backgrounds. The decay width of  $\Sigma^\pm\rightarrow W^\pm\nu$, is proportional to $\sum\limits_\alpha |V_\alpha|^2$, and therefore does not provide any information regarding the individual mixing. On the other hand,  $\Sigma^\pm \rightarrow Zl^\pm$ is sensitive to the type of mixing, and could be effectively used to identify the nature of mixing. When only one of the mixings is present, only the production is sensitive to the mixing. In case both $V_e$ as well as $V_{\mu,~\tau}$ are present, the decay becomes sensitive to all the mixings present, while the production is sensitive to $V_e$. One should in principle be able to fit the data with these parameters. Thus, the final states of $4j+2l,~~4j+l+MET$ and $2j+l+MET$ involving $\Sigma^\pm \rightarrow Zl^\pm$ could be made use of for this. At the same time, the final states of $4j+MET$, which involve only $\Sigma^\pm\rightarrow W^\pm\nu$ decay has no sensitivity to the mixing in the decay. Thus, effectively it becomes a one parameter problem to probe $V_e$. This channel also has about double, and more than four times the statistics of $4j+l+MET$ and $4j+2l$ final states, respectively. Again, a combined analysis of all the channels involved will certainly be able to constrain the mixing parameters quite efficiently. 

\begin{table}
\begin{center}
\begin{tabular}{|l|l|r|r|l|}
\hline
Final State & Signal & \multicolumn{2}{c|}{$\sigma\times$ {BR} in fb} & Background\\\cline{3-4}
&  &$V_{e}=0.05,  $& $V_{e}= 0,$&($\sigma\times$ BR in fb) \\
 &&$V_{\mu}=V_{\tau}=0$&$V_{\mu~ {\rm or}~\tau}\ne 0$& \\ 
\hline
\hline
4$j$+MET & $WW\nu\nu$& 9.36 & 10.82 & $WWZ$ (4)\\
2$j$+1$\ell$+MET&$WW\nu\nu$& 1.50 & 1.70 & $WWZ$(0.63),$WW$(64.5)\\ 
2$\ell$+MET&$WW\nu\nu$ & 0.24 & 0.26 & $WWZ$(0.1), $WW$(10.1)\\
\hline
4$j$+2$\ell$ & ZZ$\ell\ell$ & 2.54 & 2.93 &  WWZ(0.68) \\
2$j$+4$\ell$ & ZZ$\ell\ell$ & 0.12 & 0.14 &  ZZZ(0.0007) \\
2$j$+2$\ell$+MET&ZZ$\ell\ell$ & 0.72 & 0.84 & ZZZ(0.004), WWZ(0.35)\\
\hline
4$j$+1$\ell$+MET&WZ$\ell\nu$  & 4.87 & 5.64 & WWZ(2.2) \\
2$j$+3$\ell$+MET&WZ$\ell\nu$ & 0.23 & 0.27  & WWZ(0.1)\\
2$j$+2$\ell$+MET&WZ$\ell\nu$ & 0.78 & 0.89 & WWZ(0.35)\\
2$j$+1$\ell$+MET&WZ$\ell\nu$ & 1.40 & 1.60 & WWZ(0.63)\\
\hline
\end{tabular}
\caption{Final state cross-section for signal and corresponding background  for the process $e^{+}e^{-} \rightarrow \Sigma^{-}\Sigma^{+}$ with different mixing angle at the ILC with $M_{\Sigma}$=500 GeV and $\sqrt{s}$ = 2 TeV. }
\label{table:ch-finalstates}
\end{center}
\end{table}

\begin{figure}
\includegraphics[width=9cm]{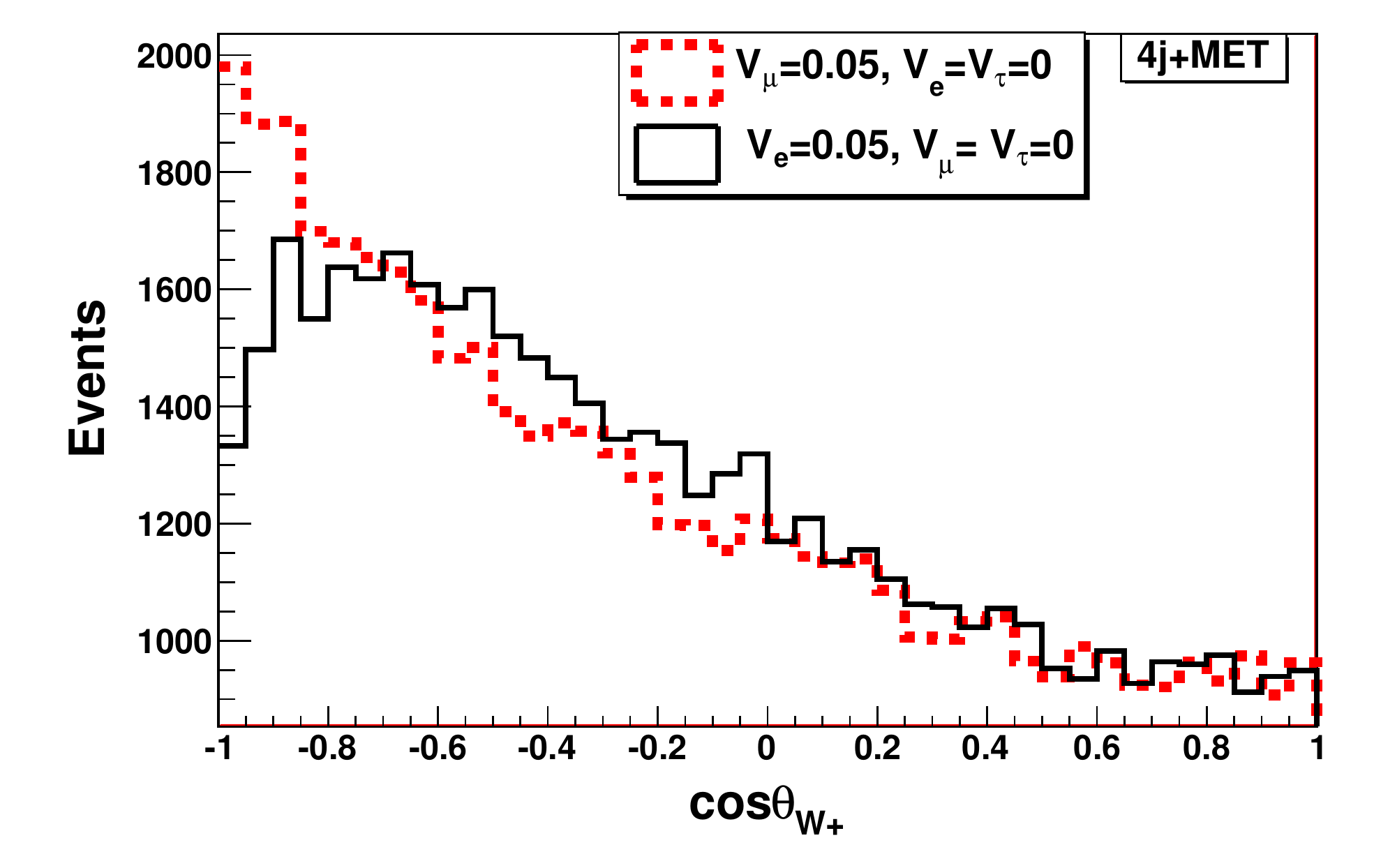}
\caption{Angular distributions of $W^{+}$ boson for final state 4j+MET of $W^{+} W^{-} \nu \nu$ signal with CM energy of 2 TeV and $M_{\Sigma}$ of 500 GeV at ILC. }
\label{fig:angW}
\end{figure}
\begin{figure}
\includegraphics[width=7.5cm]{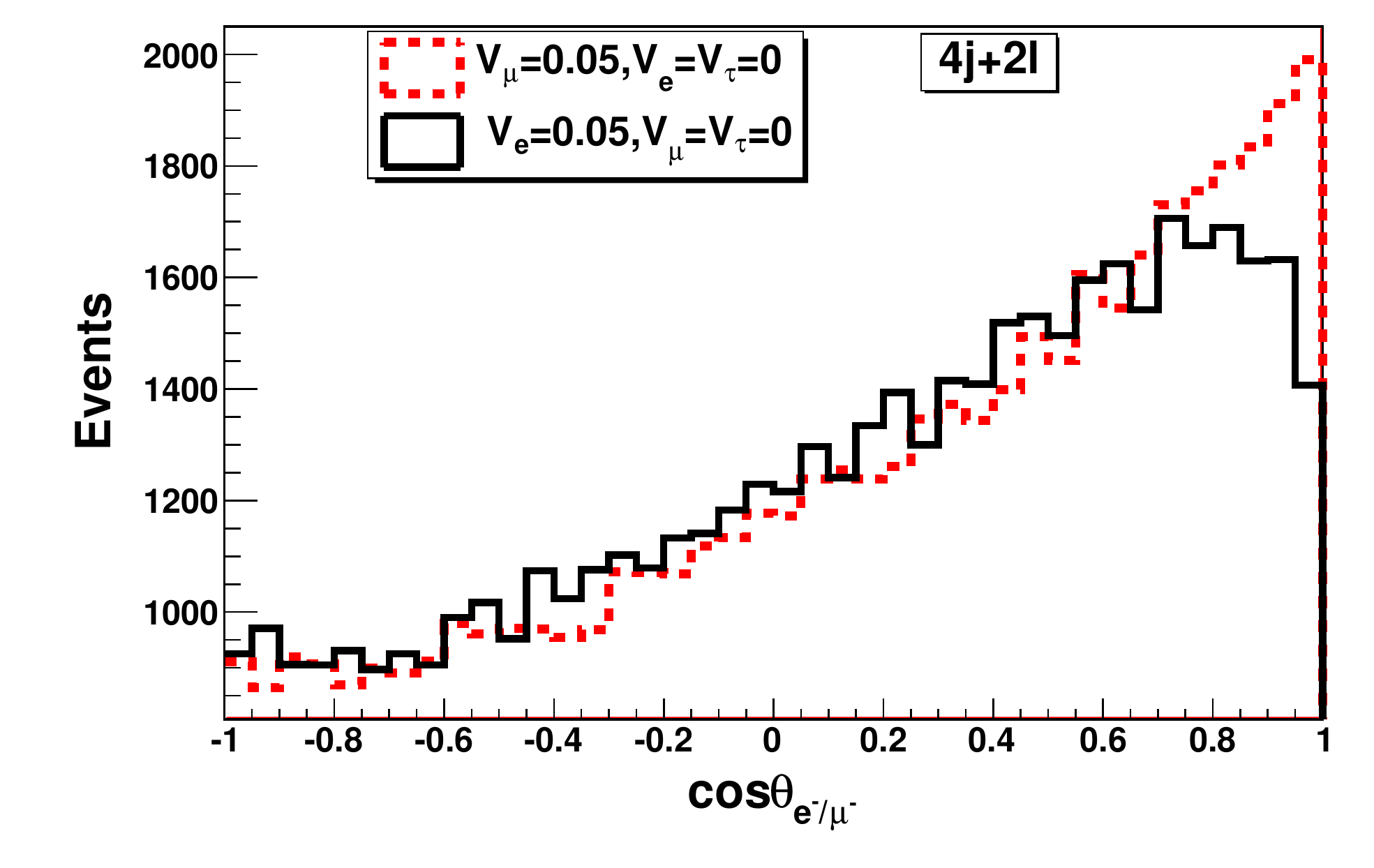}
\includegraphics[width=7.5cm]{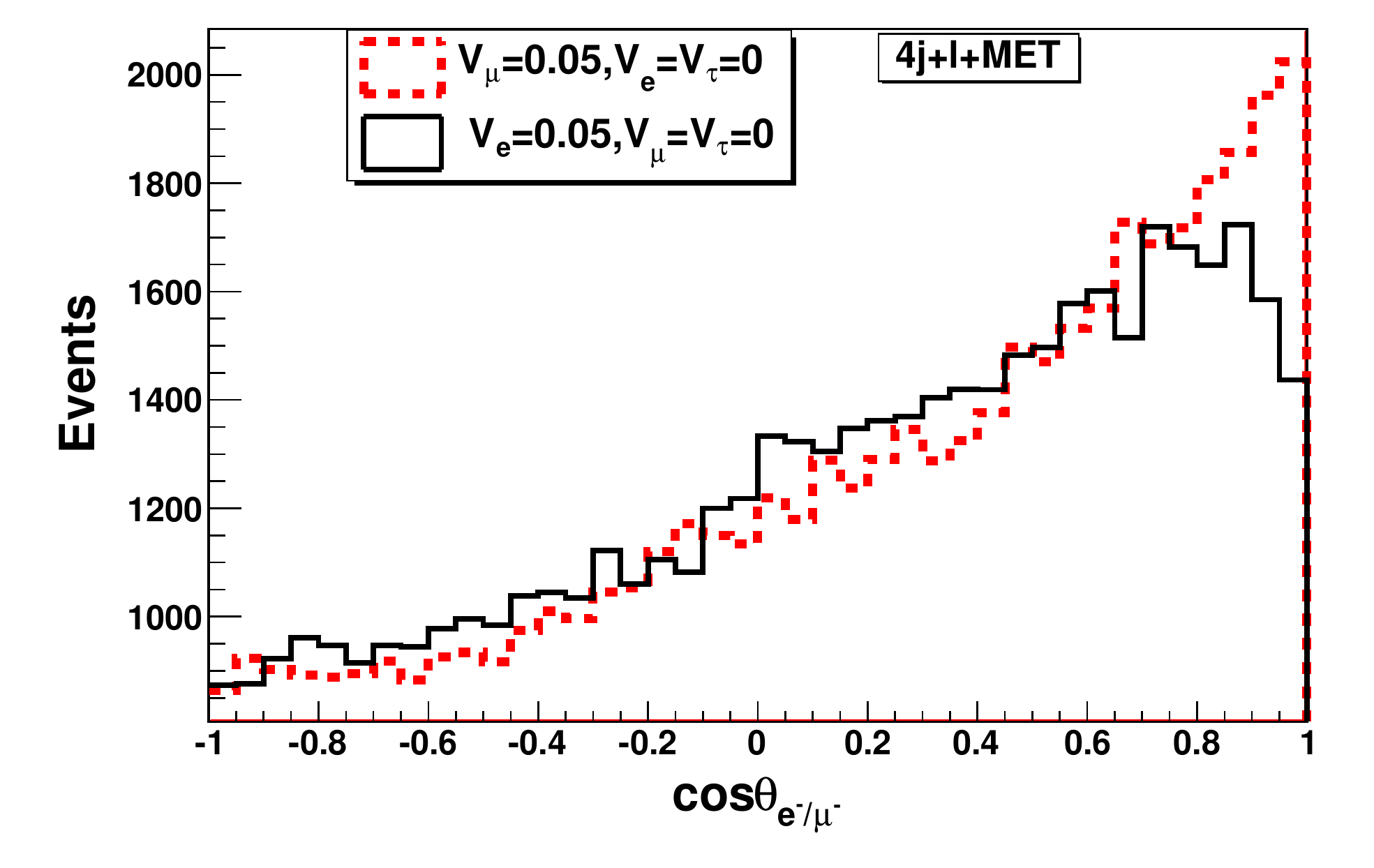}
\caption{Angular distributions of $e^{-}$(black solid) and $\mu^{-}$ (red-dashed) corresponding to the final states $4j+2\ell$ (left) and $4j+\ell$+MET (right) at $\sqrt{s}= 2$ TeV and $M_{\Sigma}= 500$ GeV at ILC. }
\label{fig:ang-l}
\end{figure}

The angular distribution of $W$ boson is shown in  Fig.~\ref{fig:angW}. The presence of the $t$-channel in the case of $V_e\ne 0$ causes the deviation in the distribution compared to the case when $V_e=0$.
These properties of the distributions of the gauge bosons are reflected in the distributions of their decay leptons. In Fig.~\ref{fig:ang-l} the angular distribution of $e^{-}$ and $\mu^{-}$  are shown corresponding to the final states $4j+2l$ and $4j+l+MET$. 

\section{Conclusions}
\label{conclusion}
The production of triplet fermions at ILC are considered. Such triplets are present in models like the type-III seesaw motivated by the small neutrino mass that such models can possibly explain. These triplet leptons can mix with the SM leptons. Direct searches at LHC limits the msses of such triplets in the range of 500 GeV, depending on different mixing scenarios. While the LHC could discover the presence of heavy triplets within a reasonable mass limit, it is not sensitive to the details of the couplings involving mixing. On the other hand, the ILC with leptonic initial states are suitable for this purpose. 

We considered the single and pair production of the neutral as well as charged triplets at an ILC. Single production is considered at a 1 TeV collider. The production is highly sensitive to the value of $V_e$, the parameter denoting the mixing of the triplet with the electrons. With a choice of $V_e=0.05$, the neutral triplet, decaying through $\Sigma^0\rightarrow Z\nu$ to a final state of $2j+$MET,  is one of the most promising channels to study this particle. The decay channel $\Sigma^0\rightarrow Wl$ resulting in a final state of $2j+l$+MET can supplement, once the SM background from $WW$ pair production is contained through analyses of the kinematic distributions like the missing energy and angular distributions. Use of beam polarisation will also be helpful as the dominant production channel is the $t$-channel with $W$ exchange, which couples only to the left-handed electrons. The single production of charged triplet also depends crucially on the value of $V_e$. The best channel to study is the $2j+2e$ resulting through the decay $\Sigma^\pm\rightarrow Ze^\pm$. Assuming $V_e=0.05$, one would be able to probe the mass up to $M_\Sigma\sim 750$ GeV. While being very sensitive to $V_e$, the single production is not sensitive to $V_\mu$ or $V_\tau$ at an $e^+e^-$ collider like ILC. A muon collider of suitable centre of mass energy and luminosity would be useful to study the mixing of the triplets with muon through single production of the heavy fermions.

Considering the pair production, we found that the neutral triplet production is quite rare, and therefore not feasible unless one considers very high luminosity. The charged triplet pair production considered at a 2 TeV ILC is promising in many ways. Firstly, the cross section is large enough when $V_e$ is absent. In fact, the presence of $V_e$ has a reducing effect on the cross section. This can be effectively used to probe the mixings. The channel with $4j+$MET final state is sensitive only to $V_e$, and thus can be used to probe this coupling. Alongside, other feasible channels like $4j+l$+MET and $4j+2l$, which are sensitive to other mixings as well,  provide a handle on different mixing scenarios.

In conclusion, although LHC could discover triplet fermions, the searches assume different mixing scenarios. Being a hadronic collider, it is hard to understand the mixings through LHC studies. The ILC, with electrons and positrons as the initial state, is suitable to study the mixing scenarios.
While the single production can be probed at a smaller centre of mass energy as long as the mixing of the triplet with electron takes reasonably large value, the pair production, though feasible only at high energy collider, is useful even when the mixings are very small. While presently it is somewhat premature, once the machine and detector details of the ILC are available, more detailed analyses will establish precise nature of the dynamics involving heavy triplet fermions.
\begin{acknowledgments}
This work is partly supported by a BRNS, DAE, Govt. of India project (2010/37P/49/BRNS/1446). The authors are highly thankful to Dr. Sumit K. Garg for useful discussions, and involvement at the initial stage of the work.
\end{acknowledgments}

\section{Appendix}
\label{appendix}
The mass basis of the Lagrangian in Eq.3 after diagonalization is given in the following Lagrangian:
\begin{equation}
\mathcal{L} =  \mathcal{L}_{Kin} + \mathcal{L}_{CC} + \mathcal{L}_{NC}^{\ell} + \mathcal{L}_{NC}^{\nu} + \mathcal{L}_{H}^{\ell} + \mathcal{L}_{H}^{\nu} + \mathcal{L}_{\eta}^{\ell} + \mathcal{L}_{\eta}^{\nu} + \mathcal{L}_{\phi^{-}}
\end{equation}
where, 
\begin{eqnarray}
\mathcal{L}_{CC} & = & \frac{g}{\sqrt{2}} \begin{pmatrix} \bar{\ell} &  \overline{\Psi}\end{pmatrix} \gamma^{\mu}W^{-}_{\mu} \left(P_{L}g_{L}^{CC} + P_{R}g_{R}^{CC}\sqrt{2} \right) \begin{pmatrix} \nu \\ \Sigma \end{pmatrix}  +h.c \\
\mathcal{L}^{\ell}_{NC} & = & \frac{g}{costh\theta_{W}} \begin{pmatrix} \overline{\ell} & \overline{\Psi}\end{pmatrix} \gamma^{\mu}Z_{\mu} \left(P_{L}g_{L}^{NC} + P_{R}g_{R}^{NC} \right) \begin{pmatrix} \ell \\ \Psi \end{pmatrix}\\
\mathcal{L}^{\nu}_{NC} & = & \frac{g}{2cos{\theta}_{W}}\begin{pmatrix}
\bar{\nu} & \overline{\Sigma^{0c}}\end{pmatrix} \gamma^{\mu}Z_{\mu}\left(P_{L}g_{\nu}^{NC}\right)
\begin{pmatrix}
\nu_{L} \\ \Sigma^{0c}
\end{pmatrix} \\
\mathcal{L}^{\ell}_{H} &=& - \begin{pmatrix}
\overline{\ell} & \overline{\Psi}\end{pmatrix}H\left(P_{L}g_{L}^{H\ell} + P_{R}g_{R}^{H\ell} \right) \begin{pmatrix}
\ell \\ \Psi \end{pmatrix}\\
\mathcal{L}_{H}^{\nu} & = & - \begin{pmatrix}\overline{\nu} & \overline{\Sigma^{0}}
\end{pmatrix}\frac{H}{\sqrt{2}}\left(P_{L}g_{L}^{H\nu} + P_{R}g_{R}^{H\nu}\right)
\begin{pmatrix}
\ell \\ \Psi \end{pmatrix}\\
\mathcal{L}_{\eta}^{\ell} & = & - \begin{pmatrix}
\overline{\ell} & \Psi \end{pmatrix} i\eta \left( P_{L}g_{L}^{\eta\ell} + P_{R}g_{R}^{\eta\ell}\right)\begin{pmatrix} \ell \\ \Psi \end{pmatrix}\\
\mathcal{L}_{\eta}^{\nu} & = & - \begin{pmatrix}
\overline{\nu} & \overline{\Sigma^{0}} \end{pmatrix}\frac{i\eta}{\sqrt{2}}\left( P_{L}g_{L}^{\eta\nu} + P_{R}g_{R}^{\eta\nu}\right)\begin{pmatrix}\nu \\\Sigma^{0} \end{pmatrix}\\
\mathcal{L}_{\phi^{-}} & = & - \begin{pmatrix}
\overline{\ell} & \overline{\Psi} \end{pmatrix}\phi^{-} \left(P_{L}g_{L}^{\phi^{-}} + P_{R}g_{R}^{\phi^{-}} \right) \begin{pmatrix}
\nu \\ \Sigma^{0} \end{pmatrix}+h.c\\
g_{L}^{CC} & = & \begin{pmatrix}
\left(1+ \frac{\epsilon}{2}\right)U_{PMNS} & -Y^{\dagger}M_{\Sigma}^{-1}\frac{\upsilon}{\sqrt{2}} \\ 0 & \sqrt{2}\left(1- \frac{\epsilon^{\prime}}{2}\right)\end{pmatrix}\\
g_{R}^{CC} & = & \begin{pmatrix}
0 & -m_{\ell}Y^{\dagger}M_{\Sigma}^{-2}\upsilon\\ -M_{\Sigma}^{-1}Y^{*}_{\Sigma}U^{*}_{PMNS}\frac{\upsilon}{\sqrt{2}} & 1- \frac{\epsilon^{\prime\star}}{2}
\end{pmatrix}\\
g_{L}^{NC} & = & \begin{pmatrix}
\frac{1}{2}- cos^{2}\theta_{W} - \epsilon & 
\frac{1}{2}Y^{\dagger}_{\Sigma}M_{\Sigma}^{-1}\upsilon \\ \frac{1}{2}M_{\Sigma}^{-1}Y_{\Sigma}\upsilon & \epsilon^{\prime}-cos^{2}\theta_{W}
\end{pmatrix}\\
g_{R}^{NC} & = & \begin{pmatrix} 1- cos^{2}\theta_{W} & m_{\ell}Y^{\dagger}_{\Sigma}M^{-2}_{\Sigma}\upsilon \\
M_{\Sigma}^{-2}Y_{\Sigma}m_{\ell}\upsilon & - cos^{2}\theta_{W}
\end{pmatrix}\\
g_{\nu}^{NC} & = & \begin{pmatrix}
1-U^{\dagger}_{PMNS}\epsilon U_{PMNS}& U^{\dagger}_{PMNS}Y^{\dagger}_{\Sigma}M^{-1}_{\Sigma}\frac{\upsilon}{\sqrt{2}} \\
\frac{\upsilon}{\sqrt{2}}M_{\Sigma}^{-1}Y_{\Sigma}U_{PMNS} & \epsilon^{\prime}
\end{pmatrix}\\
g_{L}^{H\ell} & = & \begin{pmatrix}
\frac{m_{\ell}}{\upsilon}(1-3\epsilon) & m_{\ell}Y^{\dagger}_{\Sigma}M_{\Sigma}^{-1} \\
Y_{\Sigma}(1-\epsilon) + M_{\Sigma}^{-2} Y_{\Sigma} m_{\ell}^{2} & Y_{\Sigma}Y_{\Sigma}^{\dagger}M_{\Sigma}^{-1}\upsilon \end{pmatrix}\\
g_{R}^{H\ell} & = & (g_{L}^{H\ell})^{\dagger}
\end{eqnarray}
\begin{eqnarray}
g_{L}^{H\nu} & = & \begin{pmatrix}
-\frac{\sqrt{2}}{\upsilon}U^{T}_{PMNS} m_{\nu} U_{PMNS} & U^{T}_{PMNS}m_{\nu}Y^{\dagger}_{\Sigma}M^{-1}_{\Sigma} \\
(Y_{\Sigma} - Y_{\Sigma}\frac{\epsilon}{2} - \frac{\epsilon^{\prime T}}{2}Y_{\Sigma})U_{PMNS} & Y_{\Sigma}Y^{\dagger}_{\Sigma}M_{\Sigma}^{-1}\frac{\upsilon}{\sqrt{2}}
\end{pmatrix}\nonumber \\&& = \begin{pmatrix}
-\frac{\sqrt{2}}{\upsilon}m_{\nu}^{d} & m_{\nu}^{d}U^{\dagger}_{PMNS}Y^{\dagger}_{\Sigma}M_{\Sigma}^{-1} \\
(Y_{\Sigma} - Y_{\Sigma}\frac{\epsilon}{2} - \frac{\epsilon^{\prime T}}{2}Y_{\Sigma})U_{PMNS} & Y_{\Sigma}Y^{\dagger}_{\Sigma}M_{\Sigma}^{-1}\frac{\upsilon}{\sqrt{2}}
\end{pmatrix}\\
g_{R}^{H\nu} & = & (g_{L}^{H\nu})^{\dagger} \\
g_{L}^{\eta\ell} & = & \begin{pmatrix}
-\frac{m_{\ell}}{\upsilon}(1+\epsilon) & -m_{\ell}Y_{\Sigma}^{\dagger}M^{-1}_{\Sigma} \\ Y_{\Sigma}(1-\epsilon)-M_{\Sigma}^{-2}Y_{\Sigma}m_{\ell}^{2} & \upsilon Y_{\Sigma}Y^{\dagger}_{\Sigma}M_{\Sigma}^{-1}
\end{pmatrix}\\
g_{R}^{\eta\ell} & = & -\left( g_{L}^{H\nu}\right)^{\dagger}\\
g_{L}^{\eta\ell} &= &\left( g_{L}^{\eta\ell} \right)^{\dagger}\\
g_{L}^{\eta\nu} &=& g_{L}^{H\nu}\\
g_{L}^{{\phi}^{-}} &= &\begin{pmatrix}
\sqrt{2}\frac{m_{\ell}}{\upsilon}(1-\frac{\epsilon}{2})U_{PMNS} & m_{\ell}Y^{\dagger}_{\Sigma}M_{\Sigma}^{-1} \\ \sqrt{2}m_{\ell}^{2}M_{\Sigma}^{-2}Y_{\Sigma}U_{PMNS} & 0
\end{pmatrix}\\
g_{R}^{\phi^{-}} &=& \begin{pmatrix}
-\sqrt{2}U_{PMNS}\frac{m_{\nu^{d*}}}{\upsilon} & \left[(Y^{\dagger}_{\Sigma} - \epsilon Y^{\dagger}_{\Sigma} - Y^{\dagger}_{\Sigma}\frac{\epsilon^{\prime \star}}{2} ) - 2m^{\star}_{\nu}Y^{\dagger}_{\Sigma} M_{\Sigma}^{-1} \right] \\ -\sqrt{2}Y^{*}_{\Sigma}(1-\frac{\epsilon^{\star}}{2})U_{PMNS}^{\star} & 2\left[-\frac{M_{\Sigma}}{\upsilon} \epsilon^{\prime T} + \epsilon^{\prime} \frac{M_{\Sigma}}{\upsilon} \right] 
\end{pmatrix}
\end{eqnarray}

Here $U_{PMNS}$ is the lowest order leptonic mixing matrix which is unitary. Masses of the charged leptons are given by the diagonal matrix $m_{\ell}$ with $\upsilon \equiv \sqrt{2}\langle\phi^{0} \rangle = 246$ GeV, $\epsilon = \frac{\upsilon^{2}}{2}Y^{\dagger}_{\Sigma} M_{\Sigma}^{-2}Y_{\Sigma}$, $\epsilon^{\prime} = \frac{\upsilon^{2}}{2}M_{\Sigma}^{-1}Y_{\Sigma}Y^{\dagger}_{\Sigma} M_{\Sigma}^{-1}$ and $\delta = \frac{m_{\ell}^{2}}{M_{\Sigma}^{2}}$ .

\end{document}